\documentclass{article}
\usepackage{graphicx}
\usepackage{amsmath,amssymb}
\usepackage{graphics,color,array,calc,rotating,epsfig,psfrag}
\numberwithin{equation}{section}
\usepackage{cite}
\usepackage{bm}
\usepackage{dcolumn}  
\usepackage{amsfonts}
\usepackage[mathscr]{eucal}
\def\be{\begin{equation}} \def\ee{\end{equation}}
\def\bea{\begin{eqnarray}} \def\eea{\end{eqnarray}}

\newcommand\prt{\partial}

\newcommand{\nn}{\nonumber}
\begin{document}
\baselineskip 18pt%
\begin{titlepage}
\vspace*{1mm}%
\hfill%
\vspace*{15mm}%
\hfill
\vbox{
    \halign{#\hfil         \cr
          } 
      }  
\vspace*{20mm}

\begin{center} {{\large {\bf Canonical structure of BHT massive gravity in warped $AdS_3$ sector}}}
\end{center}
\vspace*{5mm}
\begin{center}
{Davood Mahdavian Yekta \footnote{d.mahdavian@hsu.ac.ir}}\\
\vspace*{0.2cm}
{\it Department of Physics, Hakim Sabzevari University, P.O. Box 397, Sabzevar, Iran}\\
\vspace*{0.1cm}
\end{center}

\begin{abstract} 
We investigate the asymptotic structure of the three dimensional Warped Anti-de Sitter (WAdS$_3$) black holes in the Bergshoeff-Hohm-Townsend (BHT) massive gravity using the canonical Hamiltonian formalism. We define the canonical asymptotic guage generators, which produce the conserved charges and the asymptotic symmetry group for the WAdS$_3$ black holes. The attained symmetry group is described by a semi-direct sum of a Virasoro and a Ka\v{c}-Moody algebra. Using the Sugawara construction, we obtain a direct sum of two Virasoro algebras. We show that not only the asymptotic conserved charges satisfy the first law of black hole thermodynamics, but also they lead to the expected Smarr formula for the WAdS$_3$ black holes. We also show that the black hole's entropy obeys the Cardy formula of the dual conformal field theory (CFT).
\end{abstract} 

\end{titlepage}

\section{Introduction}
A theoretically physical issue in general relativity (GR) is the existence of black hole solutions, which are naturally endowed through the laws of black hole thermodynamics and accompanied by a macroscopic entropy\cite{Bardeen:1973gs,Bekenstein:1973ur}. However, we need a quantum theory of GR to produce this entropy from counting the micro-states\cite{Strominger:1996sh}. String theory is the most well-known candidate of quantum theory of GR, which embraces two practical features: the gauge/gravity duality\cite{Maldacena:1997re}, which is also called the AdS/CFT correspondence, and the black hole physics. 

Three-dimensional (3D) gravity is another candidate to study the quantum theory of GR\cite{Gott:1982qg}, which could also be appeared in the lower dimensional solutions of string theories\cite{Israel:2004vv}-\!\!\cite{Azeyanagi:2012zd}. However, 3D gravity describing by the Einstein-Hilbert action has no physical degrees of freedom\cite{Deser:1983tn,Deser:1983nh}, so it demands that we add higher derivative correction terms to the action. The first odd-parity extension is the topologically massive gravity (TMG) including a gravitational Chern-Simons Lagrangian, which gives one massive and two massless spin-2 gravitons\cite{Deser:1982vy,Deser:1981wh}. The BHT massive gravity theory is a unitary parity-preserving extension, which includes higher curvature terms up to the forth order of derivatives\cite{Bergshoeff:2009hq}. Other extensions of this theory could be found in Refs. \cite{Sinha:2010ai}-\!\!\cite{Ghodsi:2010ev}.

In this letter, we study the BHT massive gravity in the canonical first-order formalism\cite{Castellani:1981us} on the basis of WAdS$_3$ black holes, which have introduced in Refs. \cite{Anninos:2008fx} and \cite{Clement:2009gq}. A similar procedure for the BTZ black holes\cite{Banados:1992wn} has made in Ref. \cite{Blagojevic:2010ir} . In the first step, we define the gauge generators based on a constrained Hamiltonian theory. We calculate the energy and angular momentum of the WAdS$_3$ black hole from the asymptotic structure of these generators due to the gauge transformations. The Poisson bracket(PB) algebra of the asymptotic generators describe the asymptotic symmetry group (ASG) of the asymptotically warped backgrounds in the BHT massive gravity. We show that these conserved charges satisfy the black hole mechanics as the first law, the Smarr formula\cite{Smarr:1972kt}, and the Cardy formula\cite{Cardy:1986ie} in verification of the AdS/CFT correspondence. 

In the investigation of the asymptotic symmetries, we use the asymptotic boundary conditions (BCs) introduced by Comp\'{e}re and Detournay in Refs. \cite{Compere:2007in,Compere:2008cv}, just like the Brown--Henneaux BCs in Ref.\cite{Brown:1986nw}, which are introduced for the asymptotically AdS$_3$ solutions in 3D gravity theories. In the framework of guage/gravity duality, this is known as WAdS$_3$/warped CFT$_2$\cite{Song:2011sr,Detournay:2012pc}.

After having briefly discussed the canonical structure of the BHT massive gravity in Sect. 2, we will compute the asymptotic conserved charges using the asymptotic gauge generators in Sect. 3. we consider, in sect. 4, the WAdS$_3$ black hole as a solution of the BHT equations of motion and then, derive the asymptotic canonical algebra and thermodynamics. Finally, in sect. 5, we discuss about the concluding results.
\section{Canonical BHT massive gravity}
The torsion free BHT massive gravity, which, because of the new mass scale $m^2$ is known as the new massive gravity, is described by the following action\cite{Bergshoeff:2009hq} 
\be \label{LNMG} I=\frac{1}{2\kappa^2}\int d^3x \sqrt{-g} \left[R-2\Lambda_0-\frac{1}{m^2}\left(R_{\mu\nu}R^{\mu\nu}-\frac38 R^2\right)\right],\ee
where $\kappa^2\!=\!8\pi G$ is the 3D Newton's constant and $\Lambda_0$ is the cosmological constant with the dimension of ({\it mass})$^2$. Since the Lagrangian of (\ref{LNMG}) contains forth order of derivatives, the BHT gravity is an even-parity theory of massive gravitons. The variation of the action (\ref{LNMG}) under a variation $\delta g^{\mu\nu}$ yields
\be \label{eqs1} R_{\mu\nu}-\frac12 R g_{\mu\nu}+\Lambda_0 g_{\mu\nu}+\frac{1}{m^2}K_{\mu\nu}=0\,, \ee
with
\be K_{\mu\nu}=\frac34 (R_{\mu\nu}-\frac14 g_{\mu\nu} R)R-\frac14 (g_{\mu\nu}\nabla^2 -\nabla_{\mu}\nabla_{\nu})R-\nabla^2(R_{\mu\nu}-\frac12 g_{\mu\nu}R)-2(R_{\mu\nu\rho\sigma}-\frac14 g_{\mu\nu}R_{\rho\sigma})R^{\rho\sigma}.\ee

Using the canonical auxiliary fields\cite{Afshar:2014ffa}, the BHT massive gravity can be rearranged in the Chern--Simons-like formulation by the $3$-form Lagrangian,
\be\label{LNMG1} L=-\sigma\, e\cdot R+\frac{\Lambda_0}{6}\,e\cdot e\times e+h\cdot T-\frac{1}{m^2}\left(f\cdot R+\frac12\, e\cdot f\times f\right),
\ee 
where $\sigma\!=\!\pm1$ is a parameter and $e^{i}$ is the dreibein field.\footnote{The local Lorentz frame metric is $\eta_{ij}=(+,-,-)$; the Latin indices $(i,j,k,\dots)$ and the Greek indices $(\mu,\nu,\lambda,\dots)$ run over 0,1, and 2, while the letters $(\alpha,\beta,\gamma,\dots)$ run over $1,2$; $\varepsilon^{012}=1$ and $g_{\mu\nu} =\eta_{ij}{e^{i}}_{\mu} {e^{j}}_{\nu}\,.$} $\omega^{i}$ is the spin-connection and the 1-forms $f^{i}$ and $h^{i}$ are two auxiliary fields. The two 2-forms, Lorentz covariant torsion $T$ and curvature $R$, are
\be T=De=de+\omega\times e\,,\qquad R=d\omega+\frac12\, \omega\times\omega\,,\ee
and $``\,D(\omega)=d+\omega\times\,"$ is the Lorentz covariant derivative. Owing to the torsion free condition, the field $h^{i}$ acts as a Lagrange multiplier. In order to prepare the canonical Hamiltonian and primary constraints in appropriate form, we rewrite the Lagrangian (\ref{LNMG1}) as
\bea \label{LNMG2} L=\frac12\, \varepsilon^{\mu\nu\rho}\Big[-\sigma\,{e^{i}}_{\mu} R_{i\,\nu\rho}+\frac{\Lambda_0}{3}\,{\varepsilon_{ijk}}\, {e^{i}}_{\mu} {e^{j}}_{\nu} {e^{k}}_{\rho}+{h^{i}}_{\mu}  T_{i\,\nu\rho}-\frac{1}{m^2}\Big({f^{i}}_{\mu} R_{i\,\nu\rho}+\frac12\,{\varepsilon_{ijk}} {e^{i}}_{\mu} {f^{j}}_{\nu} {f^{k}}_{\rho} \Big)\Big].\eea

The equations of motion can be extracted from (\ref{LNMG1}) by variation w.r.t. $e^{i},\omega^{i},f^{i}$, and $h^{i}$, respectively,
\bea \label{eqs2}
&&-\sigma R(\omega)+\frac{\Lambda_0}{2}\, e\times e+D h-\frac{1}{2m^2}\, f\times f= 0\,,\nn\\
&&-\sigma\, T(\omega)+e\times h-\frac{1}{m^2}\, D f=0\,,\nn\\
&&\,\,\,R(\omega)+e\times f= 0\,,\\
&&\,\,\,T(\omega)=0 \,.\nn
\eea
The second and the third equations plus the torsion free condition, $T(\omega)=0$, give the following fields
\be \label{hf} 
h_{\mu\nu}=\frac{1}{m^2}\, C_{\mu\nu}\equiv\frac{1}{m^2} {\varepsilon_{(\mu|}}^{\rho\lambda}\nabla_{\rho}S_{\lambda|\nu)}\,,\quad
f_{\mu\nu}=-S_{\mu\nu}\equiv-(R_{\mu\nu}-\frac14 g_{\mu\nu} R)\,,
\ee
where $C_{\mu\nu}$ and $S_{\mu\nu}$ are the symmetric Cotton and Schouten tensors. Substituting these into the first equation of (\ref{eqs2}) and after a long calculations, it leads to the Eq. (\ref{eqs1}). 
\section{Asymptotic Conserved Charges} \label{sec3}
In the higher curvature models of gravity, the asymptotic conserved charges are generally computed from the ADM formalism\cite{Arnowitt:1962hi} for the asymptotically flat backgrounds and the AD formalism\cite{Abbott:1981ff} for the asymptotically AdS ones. In the latter case, for generic higher curvature gravity theories see Ref.\cite{Deser:2002jk}. Though linearization around an asymptotically WAdS$_3$ background have been done for TMG \cite{Bouchareb:2007yx}, we are unable to do this for the BHT gravity. Therefore we use the Hamiltonian formalism for the BHT gravity to find these conserved quantities.

The asymptotic properties are described by defining the appropriate asymptotic gauge generators by including the consistent BCs. Consequently, we find the finite conserved charges of WAdS$_3$ black holes in the BHT massive gravity. In brief, we first construct the canonical Hamiltonian from the Lagrangian (\ref{LNMG2}) using the Castellani formalism \cite{Castellani:1981us}, and then define the primary constraints.\footnote{ The necessary and sufficient conditions for $G$ as a gauge generator are \cite{Castellani:1981us}
 \be G=primary,\qquad
	 \left\{G,H\right\}=primary,\qquad
	 \left\{G,any\,constraint\right\}=constraints.\ee
The Hamilton equations yield the following primary and secondary constraints:
\be \label{psc} {G|}_{\phi_{\rho}=0}=0\,,\qquad{\{G,H_{c}\}|}_{\phi_{\rho}=0}=0\,,\ee
where $\phi_{\rho}$'s are primary constraints and $H_{c}$ is the canonical Hamiltonian of the system in the gauge theory.} 
Finally, under the local Pioncar\'{e} gauge transformations (PGTs), we calculate the asymptotic conserved charges.

We define the canonical Hamiltonian as 
\bea \label{cham}
	&& \mathcal{H}_{c}={e^{i}}_{0} \mathcal H_{i}+{\omega^{i}}_{0} \mathcal K_{i}+{f^{i}}_{0} \mathcal S_{i}+{h^{i}}_{0} \mathcal T_{i}+\prt_{\alpha} \chi^{\alpha}\,,\\
&&	\mathcal H_{i}=- \varepsilon^{0 \alpha\beta} \left(-\sigma R_{i\alpha\beta}+\Lambda_0\, \varepsilon_{ijk}\, {e^{j}}_{\alpha} {e^{k}}_{\beta}+D_{\alpha} h_{i\beta}-\frac{1}{2m^2}\varepsilon_{ijk}\, {f^{j}}_{\alpha} {f^{k}}_{\beta}\right)\,,\nn\\
&&	\mathcal K_{i}=- \varepsilon^{0 \alpha\beta} \left(-\sigma T_{i\alpha\beta}+\varepsilon_{ijk}\, {e^{j}}_{\alpha} {h^{k}}_{\beta}-\frac{1}{m^2}\,D_{\alpha} f_{i\beta}\right)\,,\nn\\
&&\mathcal S_{i}=- \varepsilon^{0 \alpha\beta}\left(-\frac{1}{m^2}R_{i\alpha\beta}-\frac{1}{m^2}\,\varepsilon_{ijk}{e^{j}}_{\alpha} {f^{k}}_{\beta}\right)\,,\nn\\
&&	\mathcal T_{i}=- \varepsilon^{0 \alpha\beta}\, T_{i\alpha\beta}\,,\nn\\
&&	\chi^{\alpha}=-\varepsilon^{0 \alpha\beta} \left({\omega^{i}}_{0}[\sigma {e_{i\beta}}+\frac{1}{m^2} f_{i\beta}]-{e^{i}}_{0}{h_{i\beta}}\right)\,,\nn
\eea
where the last term in ${\mathcal H}_{c}$ is a surface term. The canonical conjugate momenta $({\pi_{i}}^{\mu}, {\Pi_{i}}^{\mu}, {p_{i}}^{\mu},  {P_{i}}^{\mu})$ are defined for the Lagrangian variables $({e^{i}}_{\mu}, {\omega^{i}}_{\mu}, {f^{i}}_{\mu}, {h^{i}}_{\mu})$, respectively by
	\be \label{cms} {\pi_{i}}^{\mu}\equiv\frac{\prt L}{\prt {\dot {e}^{i}}_{\mu}}\,,\qquad {\Pi_{i}}^{\mu}\equiv\frac{\prt L}{\prt {\dot {\omega}^{i}}_{\mu}}\,,\qquad {p_{i}}^{\mu}\equiv\frac{\prt L}{\prt {\dot {f}^{i}}_{\mu}}\,,\qquad {P_{i}}^{\mu}\equiv\frac{\prt L}{\prt {\dot {h}^{i}}_{\mu}}\,. \ee
We need to consider the constraints in order to construct the Hamiltonian formulation for such a theory. When the conjugate momenta are not independent functions of velocities, the primary constraints contribute in the phase space. The set of primary constraints are defined from the relations (\ref{LNMG2}) and (\ref{cms})
	\bea \label{pc} &&{\phi_{i}}^{0}\equiv {\pi_{i}}^{0} \approx 0\,,\qquad {\phi_{i}}^{\alpha}\equiv {\pi_{i}}^{\alpha}-\varepsilon^{0\alpha\beta} h_{i\beta} \approx 0\,,\nn \\
	&&{\Phi_{i}}^{0}\equiv {\Pi_{i}}^{0} \approx 0\,,\qquad {\Phi_{i}}^{\alpha}\equiv {\Pi_{i}}^{\alpha}+\varepsilon^{0\alpha\beta} (\sigma e_{i\beta}-\frac{1}{m^2} f_{i\beta}) \approx 0\,,\nn\\
	&& {\psi_{i}}^{\mu}\equiv {p_{i}}^{\mu}\,\approx 0,\qquad {\Upsilon_{i}}^{\mu}\equiv {P_{i}}^{\mu}\,\approx 0 ,
	 \eea
	 and therefore the total Hamiltonian constructed out of these primary constraints becomes 
	 \be \mathcal{H}_{T}=\mathcal H_{c}+{u^{i}}_{\mu} {\phi^{\mu}}_{i}+{v^{i}}_{\mu} {\Phi^{\mu}}_{i}+{w^{i}}_{\mu} {\psi^{\mu}}_{i}+{z^{i}}_{\mu} {\Upsilon^{\mu}}_{i}\,.\ee 
The consistency conditions accompanied with the primary constraints (\ref{pc}) give a number of the secondary constraints, which are extensively discussed in Ref.\cite{Blagojevic:2010ir}. The secondary constraints, in fact, come out as the result of
the consistency of the primary constraints in Eq.(\ref{psc}). The canonical structure of the asymptotic symmetry is describing by the following canonical gauge generators
	 \bea \label{GG}
	 G\!\!\!\!&=&\!\!\!-G_1-G_2\,,\\
	 G_{1}\!\!\!\!&=&\!\!\! \dot\xi^{\rho}\left({e^{i}}_{\rho}{\pi_{i}}^{0}+{\omega^{i}}_{\rho}{\Pi_{i}}^{0}+{f^{i}}_{\rho}{p_{i}}^{0}+{h^{i}}_{\rho}{P_{i}}^{0}\right)\nn\\
	 &&+\,\xi^{\rho}\left[ {e^{i}}_{\rho} \bar{\mathcal {H}}_{i}+{\omega^{i}}_{\rho} \bar{\mathcal {K}}_{i}+{f^{i}}_{\rho} \bar{\mathcal {S}}_{i}+{h^{i}}_{\rho} \bar{\mathcal {T}}_{i}+(\prt_{\rho} {e^{i}}_{0}) {\pi_{i}}^{0}+(\prt_{\rho} {\omega^{i}}_{0}) {\Pi_{i}}^{0}+(\prt_{\rho} {f^{i}}_{0}) {p_{i}}^{0}+(\prt_{\rho} {h^{i}}_{0}) {P_{i}}^{0}\right]\,,\nn\\
	 G_2\!\!\!\!&=&\!\!\! \dot{\theta}^{i}{\Pi_{i}}^{0}+\theta^{i}\left[\bar{\mathcal K}_{i}-\varepsilon_{ijk}\left({e^{j}}_{0}{\pi}^{k0}+{\omega^{j}}_{0}{\Pi}^{k0}+{f^{j}}_{0}{p}^{k0}+{h^{j}}_{0}{P}^{k0}\right)\right]\,,\nn
	 \eea
and the local PGTs are 
	 \bea \label{PGT}
	 \delta_{0}{e^{i}_{\mu}}\!\!\!&=&\!\!\!-{\varepsilon^{i}}_{jk} {e^{j}}_{\mu} \theta^{k}-(\prt_{\mu} \xi^{\rho}){e^{i}}_{\rho}-\xi^{\rho} \prt_{\rho}{e^{i}}_{\mu}\,,\nn\\	 \delta_{0}{\omega^{i}_{\mu}}\!\!\!&=&\!\!\!-\nabla_{\mu}\theta^{i}-(\prt_{\mu} \xi^{\rho}){\omega^{i}}_{\rho}-\xi^{\rho} \prt_{\rho}{\omega^{i}}_{\mu}\,,\nn\\
	 \delta_{0}{f^{i}_{\mu}}\!\!\!&=&\!\!\!-{\varepsilon^{i}}_{jk} {f^{j}}_{\mu} \theta^{k}-(\prt_{\mu} \xi^{\rho}){f^{i}}_{\rho}-\xi^{\rho} \prt_{\rho}{f^{i}}_{\mu}\,,\\
	 \delta_{0}{h^{i}_{\mu}}\!\!\!&=&\!\!\!-{\varepsilon^{i}}_{jk} {h^{j}}_{\mu} \theta^{k}-(\prt_{\mu} \xi^{\rho}){h^{i}}_{\rho}-\xi^{\rho} \prt_{\rho}{h^{i}}_{\mu}\,.\nn
\eea  
Here, the gauge symmetries are the asymptotic local translations $\xi^{\mu}$ and the local Lorentz rotations $\theta^{i}$ of Poincar\'{e} transformations. Due to the BCs given in the next section and the asymptotic behavior of $\theta^{i}$, the variation of $G_2$ term vanishes after the integration. Varying the $G_1$ term, we have from (\ref{cham})
\bea \label{delg1}
\delta G_1\!\!\!\!&=&\!\!\!\!\ \xi^{\rho} \left( {e^{i}}_{\rho} \,\delta{\mathcal H_{i}}+{\omega^{i}}_{\rho} \,\delta{\mathcal K_{i}}+{f^{i}}_{\rho} \,\delta{\mathcal S_{i}}+{h^{i}}_{\rho} \,\delta{\mathcal T_{i}}\,\right)+\prt \mathcal O_1+R\\
\!\!\!\!&=&\!\!\!\!2 \,\varepsilon^{0\alpha\beta}\,\xi^{\rho}\prt_{\alpha}\Big[{e^{i}}_{\rho}\,(\sigma\, \delta \omega_{i\beta}-\frac12\, \delta h_{i\beta})+{\omega^{i}}_{\rho}\,(\sigma \,\delta e_{i\beta}+\,\frac{1}{2\,m^2}\,\delta f_{i\beta})+\frac{1}{m^2} {f_{i}}_{\rho} \delta \omega_{i\beta}-{h^{i}}_{\rho}\, \delta e_{i\beta}\Big],\nn
\eea 
where $\prt \mathcal O_1$ is a boundary term that vanishes after integration and $R$ includes the regular terms. The expression $\mathcal O_{n}$ stands for $\mathcal O (r^{-n})$, and from the Stoke's theorem
\be \label{ST} \int_{\mathcal M_2} d^2 x\prt_{\alpha}v^{\alpha}=\int_{\prt \mathcal M_2} v^{\alpha} df_{\alpha}=\int_{0}^{2\pi} v^1 d\varphi\qquad (df_{\alpha}=\varepsilon_{\alpha\beta}dx^{\beta})\,,\ee
we can show the term $\prt \mathcal O_1$ has no contribution to the asymptotic conserved charges, by the fact that the boundary of ${\mathcal M_2}$ is a circle at infinity parametrized by the angular coordinate $\varphi$.

We can arrange the relation (\ref{delg1}) to the following form 
\be
\delta G_1=\prt_{\alpha} (\xi^{0} \delta \mathcal E^{\alpha}+\xi^{2}\delta \mathcal M^{\alpha})\,,
\ee
where 
\bea \label{eam}
\mathcal E^{\alpha}\!\!\!\!\!\!&=&\!\!\!\!2\, \varepsilon^{0\alpha\beta}\left[{e^{i}}_{0}\,(\sigma\, \delta \omega_{i\beta}-\frac12\, \delta h_{i\beta})+{\omega^{i}}_{0}\,(\sigma \,\delta e_{i\beta}+\frac{1}{2\,m^2}\delta f_{i\beta})+\frac{1}{m^2} {f^{i}}_0 \delta \omega_{i\beta}-{h^{i}}_{0}\, \delta e_{i\beta}\right]\!,\nn\\
\mathcal M^{\alpha}\!\!\!\!\!\!&=&\!\!\!\!2\,\varepsilon^{0\alpha\beta}\left[{e^{i}}_{2}\,(\sigma\, \delta \omega_{i\beta}-\frac12\, \delta h_{i\beta})+{\omega^{i}}_{2}\,(\sigma \,\delta e_{i\beta}+\frac{1}{2\,m^2}\delta f_{i\beta})+\frac{1}{m^2} {f^{i}}_2 \delta \omega_{i\beta}-{h^{i}}_{2}\, \delta e_{i\beta}\right]\!.
\eea
The energy and angular momentum of the black hole are the conserved charges corresponding to the diffeomorphisms $\xi^{0}=1$ and $\xi^{2}=1$, respectively,
\be \label{eam2} E=\int_{0}^{2\pi}   \mathcal E^1\,d\varphi\quad,\quad J=\int_{0}^{2\pi}   \mathcal M^1\,d\varphi\,.\ee

\section{WAdS$_3$ black hole}
In this section, we investigate the structure of the BHT fields content in the canonical form on the basis of WAdS$_3$ black hole. The line element of this solution is given by the metric\cite{Anninos:2008fx} 
\be \label{wads1} ds^2=N^2 dt^2-\frac{l^2 dr^2}{4 N^2 K^2}-K^2(d\varphi+N_{\varphi} dt)^2\,,\ee
where the functions $N, K$, and $N_{\varphi}$ are
\bea \label{wads} N^2=\frac{(\nu^2+3)(r-r_{+})(r-r_{-})}{4K^2}\,,\quad N_{\varphi}=\frac{2\nu r-\sqrt{ (\nu^2+3)\,r_{+} r_{-}}}{2 K^2}\,,\nn\\
K^2=\frac{r}{4}\left[3(\nu^2 -1)+(\nu^2+3)(r_{+}+ r_{-})-4\nu \sqrt{ (\nu^2+3)\,r_{+} r_{-}}\right]\,.\eea
The parameter $\nu$ is a warped factor and $r_{+}$, $r_{-}$ are the outer and inner horizons of the black hole, respectively. Substituting (\ref{wads1}) into the equations of motion (\ref{eqs1}), we obtain 
\be \label{wval} l^2=-\frac{4\nu^4-48\nu^2+9}{2\,\Lambda_0 (20\nu^2-3)}\,,\qquad m^2=\frac{20\nu^2-3}{2\,l^2}.\ee
As we see, the warped factor $\nu$ corrects the radius of AdS space from $l^2=-1/\Lambda_0$ to the above value in the presence of the higher order corrections. The components of the diagonal dreibein $e^{i}$ for  the metric (\ref{wads1}) are
\be \label{wads2} e^{0}=N dt\,,\quad e^{1}=\frac{l}{2NK} dr\,,\quad e^{2}=K(d\varphi+N_{\varphi} dt)\,,\ee
and the components of the spin connection are computed from the torsion-free condition
\be \label{spc} \omega^{0}=-\frac{N\nu}{l} dt-\frac{2NKK'}{l}d\varphi\,,\quad \omega^{1}=-\frac{K N_{\varphi}'}{2N} dr\,,\quad \omega^{2}=-\frac{KN_{\varphi}\nu}{l}\, dt+\frac{K^3 N_{\varphi}'}{2N} d\varphi\,.\ee
Inserting the fields (\ref{wads2}) and (\ref{spc}) into the second and third equations in (\ref{eqs2}), we arrive at 
\bea \label{ff}
f_0\!\!\!&=&\!\!\!-l^{-2}\left[(2\nu^2-\frac32) N dt+3(\nu^2-1) N K^2 N_{\varphi} d\varphi \right]\,,\nn\\
f_1\!\!\!&=&\!\!\!-l^{-2}\,(\nu^2-\frac32)\,\frac{l}{2 N K}\,dr\,,\\
f_2\!\!\!&=&\!\!\!-l^{-2}\,\left[ (\frac32-2\nu^2)K N_{\varphi}\,dt+\left(\frac32-2\nu^2-3(\nu^2 -1)N^2\right)\,K d\varphi\right]\,,\nn\eea
and 
\bea \label{hfs}
h_0\!\!\!&=&\!\!\!\frac{3\nu}{m^2 l^3} (\nu^2-1) N (2+3N^2)\,dt\,,\nn\\
h_1\!\!\!&=&\!\!\!\frac{3\nu}{m^2 l^3} (\nu^2-1) \frac{3 N_{\varphi} l}{2}\, dr\,,\\
h_2\!\!\!&=&\!\!\!-\frac{3\nu}{m^2 l^3} (\nu^2-1)\left[3 N^2 K N_{\varphi}\, dt- 2 K d\varphi\right].\nn
\eea
Now, we have all the fields content of the theory (\ref{LNMG1}) and ready to compute the conserved charges of the WAdS$_3$ black hole.
\subsection{Conserved Charges}
The ASG of asymptotically WAdS$_3$ metrics is described by four Killing vectors relating to the isometry group $SL(2,R)\times U(1)$ \cite{Anninos:2008fx}. In fact, the asymptotic symmetries are defined by the transformations that do not affect the asymptotic form of the field configurations. Having well-defined canonical generators in the ASG, we must introduce suitable BCs. 
For the asymptotically WAdS-like spacetimes, such as G\"{o}del and spacelike stretched AdS$_3$ black holes, these BCs have been proposed in Refs. \cite{Compere:2007in,Compere:2008cv} as follows
\be \label{wbc1}
{g}_{\mu\nu}={\bar g}_{\mu\nu}+{G}_{\mu\nu},\,\, {\bar g}_{\mu\nu}=\!\!\left(\begin{array}{ccc}
 -1 \!\!&\!\! 0 \!\!&\!\! -\nu r \\ 
0 \!\!&\!\! -\frac{l^2}{(\nu^2 +3)r^2} \!\!&\!\! 0 \\ 
-\nu r \!\!&\!\! 0 \!\!&\!\! -\frac34(\nu^2 -1)r^2 
\end{array} \right)\!\!,\,
{G}_{\mu\nu}\sim\left(\begin{array}{ccc}
 \mathcal O_{1} & \mathcal O_{2} & \mathcal O_{0} \\ 
\mathcal O_{2} & \mathcal O_{3} & \mathcal O_{1} \\ 
\mathcal O_{0} & \mathcal O_{1} & \mathcal O_{-1} 
\end{array} \right)\!.
\ee 

We can use this asymptotic form of the metric to derive the asymptotic behavior of the triad fields\cite{Blagojevic:2009ek}. So, from the relations (\ref{wads2})-(\ref{hfs}) and according to (\ref{wbc1}), we obtain
\be \label{bc1} {e^{i}}_{\mu}\!=\!{\bar e^{i}}_{\mu}+{E^{i}}_{\mu},\quad {\omega^{i}}_{\mu}\!=\!{\bar\omega^{i}}_{\mu}+{\Omega^{i}}_{\mu},\quad {f^{i}}_{\mu}\!=\!{\bar f^{i}}_{\mu}+{F^{i}}_{\mu},\quad {h^{i}}_{\mu}\!=\!{\bar h^{i}}_{\mu}+{H^{i}}_{\mu}\,,\ee
where the sub-leading boundary terms are
\bea \label{bc2} &&{E^{i}}_{\mu}\!\sim\!\left(\begin{array}{ccc}
 \mathcal O_{1} & \mathcal O_{2} & \mathcal O_{2} \\ 
\mathcal O_{2} & \mathcal O_{2} & \mathcal O_{1} \\ 
\mathcal O_{1} & \mathcal O_{2} & \mathcal O_{0} 
\end{array} \right),\quad
{\Omega^{i}}_{\mu}\!\sim\!\left(\begin{array}{ccc}
 \mathcal O_{1} & \mathcal O_{2} & \mathcal O_{0} \\ 
\mathcal O_{2} & \mathcal O_{2} & \mathcal O_{1} \\ 
\mathcal O_{1} & \mathcal O_{2} & \mathcal O_{0} 
\end{array} \right),\nn\\
 &&{F^{i}}_{\mu}\!\sim\!\left(\begin{array}{ccc}
 \mathcal O_{1} & \mathcal O_{2} & \mathcal O_{0} \\ 
\mathcal O_{2} & \mathcal O_{2} & \mathcal O_{1} \\ 
\mathcal O_{1} & \mathcal O_{2} & \mathcal O_{0} 
\end{array} \right),\quad
{H^{i}}_{\mu}\!\sim\!\left(\begin{array}{ccc}
 \mathcal O_{1} & \mathcal O_{2} & \mathcal O_{0} \\ 
\mathcal O_{2} & \mathcal O_{2} & \mathcal O_{1} \\ 
\mathcal O_{1} & \mathcal O_{2} & \mathcal O_{0} 
\end{array} \right)\,.
\eea
Note that the bared notation in (\ref{bc1}) refers to the leading-order of the background fields and the asymptotic configurations should include the warped black hole geometries. The subset of the PGTs (\ref{PGT}), which leaves the BCs (\ref{bc1}) and (\ref{bc2}) invariant, gives the following vectors
\bea\label{diff}
\xi^{0}\!\!\!&=&\!\!\! T(\varphi)+\mathcal O_2\,,\quad \theta^0=-\frac{2l}{\sqrt{3(\nu^2+3)(\nu^2-1)}\,r}\,\prt_2^2 S(\varphi)+\mathcal O_2,\,\nn\\
\xi^{1}\!\!\!&=&\!\!\!-r \prt_2 S(\varphi) +\mathcal O_{0}\,,\quad
\theta^1=\frac{2l\sqrt{\nu^2+3}}{3(\nu^2-1)\,r}\,\prt_2 T(\varphi)+\mathcal O_3,\,\\
\xi^{2}\!\!\!&=&\!\!\!S(\varphi)+\mathcal O_2\,,\quad \theta^2=-\frac{4l\nu}{(\nu^2+3)\sqrt{3(\nu^2-1)}\,r}\,\prt_2^2 S(\varphi)+\mathcal O_2\,,\nn
\eea
where the functions $T(\varphi)$ and $S(\varphi)$ are some harmonic functions of the periodic coordinate $\varphi$. The PGTs produce a closed Lie algebra,
\be \label{pbpgt}[\delta'_0,\delta ''_0]=\delta'''_0(T''',S''')\,,\ee
such that to lowest order, we obtain
\be \label{cr} 
T'''\!=\!S'\prt_2T''-S''\prt_2 T',\quad S'''\!=\!S'\prt_2S''-S''\prt_2 S'\,.
\ee

The improved form of the gauge generator (\ref{GG}) is $\tilde{G}=G+K$\cite{Blagojevic:2009ek}, where the surface boundary term 
\be \label{bst} K=\oint df_{\alpha} \left(\xi^{0} \mathcal E^{\alpha}+\xi ^2 \mathcal M^{\alpha}\right)={\int_{0}}^{2\pi} d\varphi (l T \mathcal E^1+S \mathcal M^1)\,,\ee
depends only on the leading terms of the gauge symmetries and $\mathcal E^1$ and $\mathcal M^1$, which are given by the expressions
\bea \label{weam1}
\mathcal E^1\!\!\!\!\!\!&=&\!\!\!\!2\,\left[{e^{i}}_{0}\,(\sigma\, \delta \omega_{i 2}-\frac12\, \delta h_{i 2})+{\omega^{i}}_{0}\,(\sigma \,\delta e_{i 2}+\frac{1}{2\,m^2}\delta f_{i 2})+\frac{1}{m^2} {f^{i}}_0 \delta \omega_{i 2}-{h^{i}}_{0}\, \delta e_{i 2}\right]\!\!,\nn\\
\mathcal M^1\!\!\!\!\!\!&=&\!\!\!\!2\,\left[{e^{i}}_{2}\,(\sigma\, \delta \omega_{i 2}-\frac12\, \delta h_{i 2})+{\omega^{i}}_{2}\,(\sigma \,\delta e_{i 2}+\frac{1}{2\,m^2}\delta f_{i 2})+\frac{1}{m^2} {f^{i}}_2 \delta \omega_{i 2}-{h^{i}}_{2}\, \delta e_{i 2}\right]\!\!.
\eea
From the adopted asymptotic BCs (\ref{bc1}) and (\ref{bc2}), the energy and angular momentum of the WAdS$_3$ black hole (\ref{eam}) have the finite values
\bea \label{weam}
E\!\!\!&=&\!\!\!\frac{\nu^2 (\nu^2+3)}{G (20\nu^2-3) }\left[r_{+}+r_{-}-\frac{1}{\nu}\sqrt{(\nu^2+3)r_{+}r_{-}}\right],\nn\\ 
J\!\!\!&=&\!\!\!\frac{\nu^3 (\nu^2+3)l}{4G (20\nu^2-3) }\left[\left(r_{+}+r_{-}-\frac{1}{\nu}\sqrt{(\nu^2+3)r_{+}r_{-}}\right)^2-(r_{+}-r_{-})^2\right].
\eea
We have multiplied the coefficient $\frac{1}{2\kappa^2}$, which, for simplicity, have been dropped in generators (\ref{GG}). The results are different in some coefficients by those in Ref.\cite{Giribet:2012rr}, which this may refer to the notations.
\subsection{Asymptotic canonical algebra}
The PB algebra of the improved asymptotic generators forms a centrally extended representation of the ASG. On the other words, though the PB algebra of the generators $\tilde{G}[\xi]$ is isomorphic to the Lie algebra of the asymptotic symmetries by (\ref{pbpgt}), it includes a centrally extended term. That is, for $\tilde{G}'\equiv \tilde{G}[T',S']$ and $\tilde{G}''\equiv \tilde{G}[T'',S'']$, we obtain 
\be \label{PB}\{\tilde{G}'', \tilde{G}'\}= \tilde{G}'''+k\,,\ee
where the functions $T''',S'''$ are given by (\ref{cr}) and $k$ is called the {\it central charge} of algebra.

In the canonical algebra, when the constraints remain unchanged under the gauge transformations $\delta_0$, we can approximate the PB as $\{\tilde{G}'', \tilde{G}'\}= \delta'_0\tilde{G}''\approx \delta'_0 K''$. So, for $\tilde{G}'''\approx K'''$, we have $\label{pc2} \delta'_0 K''\approx K'''+k$.
Using the PGTs laws, the expression for $\delta'_0 K''$ reduces to 
\bea \label{pbst}
\delta_0  \mathcal E^{1}&=&-S\prt_2  \mathcal E^1-(\prt_2 S) \mathcal E^1-\frac{8 \nu(\nu^2+3)}{3(20\nu^2-3)}\,\prt_2 T\,,\\
\delta_0  \mathcal M^{1}&=&-2(\prt_2 S) \mathcal M^1-S \prt_2  \mathcal M^1-(l\prt_2 T)\mathcal E^1-\frac{64\nu^3 l}{(\nu^2+3)(20\nu^2-3)}\,\prt_2^3 S,\nn
\eea
such that from the Eqs. (\ref{cr}) and (\ref{bst}), we obtain
\be \label{ccint} k=\frac{8 \nu l(\nu^2+3)}{3(20\nu^2-3)} \int_{0}^{2\pi}d\varphi T'' \prt_2 T'+\frac{64\nu^3 l}{(\nu^2+3)(20\nu^2-3)}\int_{0}^{2\pi}d\varphi S''\prt_2^3 S'.\ee

By defining two sets of the Fourier modes from the canonical generators
\be \mathcal{P}_{n}\equiv\tilde{G}(T=e^{-i n \varphi},S=0)\,,\quad \mathcal L_{n}\equiv\tilde{G}(T=0\,, S=e^{-i n \varphi})\,,\ee
the canonical algebra takes the form 
\bea \label{calg}
i\{\mathcal L_{n},\mathcal L_{m}\}\!\!\!&=&\!\!\!(n-m) \mathcal L_{n+m}+ \frac{c_{V}}{12}\, n^3 \delta_{n+m,0}\,,\nn\\
i\{\mathcal P_{n},\mathcal P_{m}\}\!\!\!&=&\!\!\!-\frac{c_{K}}{12} n \delta_{n+m,0}\,,\nn\\
i\{\mathcal P_{n},\mathcal L_{m}\}\!\!\!&=&\!\!\!n \mathcal P_{n+m}\,,
\eea
where 
\be \label{cc1}
c_{V}=\frac{96 \nu^3 l}{G(\nu^2+3)(20\nu^2-3)}\,,\qquad c_{K}=\frac{4\nu l(\nu^2+3)}{G (20\nu^2-3)}\,,
\ee
are the central extended terms of a Virasoro and a Ka\v{c}-Moody algebra, respectively. We use the Sugawara construction \cite{Sugawara:1967rw} to produce the conformal algebra for the basis of (\ref{calg}). We define the following set of generators 
\be \label{newgens} L_{n}\equiv-\frac{6}{c_{k}}\sum_{r}\mathcal P_{r} \mathcal P_{n-r}\Rightarrow \qquad L_{n}^{-}\equiv \mathcal L_{n}-L_{n}\quad,\quad L_{n}^{+}\equiv -L_{-n}-i n z \mathcal P_{-n},\ee 
whereupon, the PB of $L_{n}^{\pm}$ take the familiar conformal form of two Virasoro algebras
\bea
i\{L_{n}^{+}, L_{m}^{+}\}\!\!\!&=&\!\!\!(n-m) L_{n+m}^{+}+ \frac{c^{+}}{12}\, n^3 \delta_{n+m,0}\,,\nn\\
i\{L_{n}^{+}, L_{m}^{-}\}\!\!\!&=&\!\!\!0\,,\nn\\
i\{L_{n}^{-}, L_{m}^{-}\}\!\!\!&=&\!\!\!(n-m) L_{n+m}^{-}+ \frac{c^{-}}{12}\, n^3 \delta_{n+m,0}\,,
\eea
with $c^{-}\!\!=\!\!c_{V}$ and $c^{+}\!\!=\!\!c_{K} z^2$. Since the BHT massive gravity is an even-parity theory of higher derivative gravity in 3D, the content of the left-right sectors in the dual CFT$_2$ are equivalent, i.e., $c^{+}=c^{-}$ and consequently we obtain
\be z^2=\frac{24\nu^2}{(\nu^2+3)^2}\,.\ee  

\subsection{Thermodynamics}
In 3D massive gravity theories, due to the higher derivative corrections, the entropy of black holes is not precisely equal to the value $\frac{Area(H)}{4G}$ proposing by Bekenstein and Hawking\cite{Bardeen:1973gs,Bekenstein:1973ur}. In general, we use the Wald formula in the case of higher Reimannian curvature terms\cite{Wald:1993nt,Iyer:1994ys}. According to the AdS/CFT duality, the macroscopic entropy of a black hole must agree with the microscopic one given by the Cardy formula, which counts the micro-states in the dual CFT. This entropy together with the other physical parameters of the black hole, such as the energy and angular momentum, satisfy the first law of black hole thermodynamics. The first law can also be obtained from the Smarr's method. For the WAdS$_3$ black holes this formula is\cite{Moussa:2008sj}
\be \label{smf} E=T_{H} S_{BH}+2\Omega_{H} J\,.\ee

Since we have computed the energy and angular momentum from the Hamiltonian formalism, we can use the Smarr-like formula (\ref{smf}) to find the entropy. The Hawking temperature and angular velocity of the event horizon  for (\ref{wads1}) are given in Ref.\cite{Anninos:2008fx} by
\be \label{TV} T_{H}=\frac{\nu^2+3}{4\pi l}\,\frac{r_{+}-r_{-}}{2\, r_{h}}\,,\qquad \Omega_{H}=\frac{1}{l\, r_{h}}\,,\quad r_{h}\equiv\nu\, r_{+}-\frac12 \sqrt{(\nu^2 +3)\,r_{+}r_{-}}\,.\ee
Substituting the values of (\ref{weam}) and (\ref{TV}) in the relation (\ref{smf}), we have 
\be \label{ent} S_{BH}=\frac{2\pi r_{h}}{4G}\left(\frac{16 \nu^2 l}{20\nu^2 -3}\right).\ee
Recently, the authors of\cite{Detournay:2016gao} have also computed the entropy of WAdS$_3$ black hole in the covariant phase space formalism, which is equal to (\ref{ent}). We have discussed the asymptotic conformal structure of WAdS$_3$ in the previous subsection, so, we can compute the entropy from the holographic consideration by the Cardy formula
\be \label{cardy1} S=\frac{\pi^2 l}{3}(c_{L} T_{L}+c_{R} T_{R})\,,\ee
where the left and right temperatures for the WAdS$_3$ black hole are \cite{Anninos:2008fx} 
\be T_{L}\equiv\frac{\nu^2+3}{8\pi l} (r_{+}-r_{-}),\quad T_{R}\equiv\frac{\nu^2+3}{8\pi l}\left(r_{+}+r_{-}-\frac{1}{\nu}\sqrt{(\nu^2+3)\,r_{+}r_{-}}\right).\ee
By using the value of central charges $c_{L}=c_{R}=c_{V}$ in (\ref{cc1}), we obtain again the entropy (\ref{ent}). We can define the following zero modes for the Fourier modes (\ref{newgens}), as we have done in Ref.\cite{Ghodsi:2011ua},
\bea L_0^{+}\!\!\!&\equiv&\!\!\!\frac{G (20\nu^2-3)}{4\nu(\nu^2 +3)} E^2=\frac{\nu^3 (\nu^2 +3)}{4G(20\nu^2 -3)} \left(r_{+}+r_{-}-\frac{1}{\nu}\sqrt{(\nu^2 +3)\,r_{+} r_{-}}\right)^2=\frac{\pi^2 l}{6} \,c_{L} T_{L}^2\,,\nn\\
L_0^{-}\!\!&\equiv&\!\!L_0^{+}-J/l=\frac{\nu^3 (\nu^2+3)}{4G(20\nu^2-3)}(r_{+}-r_{-})^2=\frac{\pi^2 l}{6}\, c_{R} T_{R}^2\,,\eea
which again by using the other Cardy formula
\be \label{cardy2} S=2\pi\left(\sqrt{\frac{c^{+} L_{0}^{+}}{6}}+\sqrt{\frac{c^{-} L_{0}^{-}}{6}}\right), \ee
and $c^{+}=c^{-}=c_{V}$, the result is in agreement with (\ref{ent}).
\section{Conclusions}
In this paper, we have studied the WAdS$_3$ black hole in the 3D BHT massive gravity and found its properties in the view point of the AdS/CFT correspondence. Since we are unable to linearize the BHT theory around the asymptotically warped backgrounds, we should try another approach. Therefore, we have studied it in the first-order formalism by constructing the canonical Hamiltonian and derived the field equations of motion. Then, we have found the conjugate momenta of the 1-form fields, which they determined the primary constraints in Eq. (\ref{pc}). We have also defined the canonical gauge generators (\ref{GG}), which under the PGTs (\ref{PGT}), lead to the finite asymptotic conserved charges (\ref{eam}). 

To illustrate the procedure, we calculated these conserved charges for the WAdS$_3$ black hole, as energy and angular momentum by (\ref{weam}). It must be noted that these values are achieved by defining suitable BCs (\ref{bc1}) and (\ref{bc2}) in the asymptotic region. We have derived the ASG of the asymptotically warped solutions using the PB of improved gauge generators, which has the $SL(2,R)\times U(1)$  isometry, and show that from the Sugawara construction it enhances to the $SL(2,R)\times SL(2,R)$  isometry with equal left-right central charges of the value $c_{V}$ in (\ref{cc1}). Finally, we calculated the entropy of the black hole from the Smarr formula and considered that the physical quantities satisfy the first law of black hole thermodynamics, $dM=T dS+\Omega dJ$. We have shown that the value of entropy (\ref{ent}) was precisely equal to the ones obtained from the Cardy formulae (\ref{cardy1}) and (\ref{cardy2}). 
As a practical matter, we can apply similar calculation for the extended theories of 3D massive gravities posed in Refs.\cite{Sinha:2010ai}-\!\!\cite{Ghodsi:2010ev} and \cite{Afshar:2014ffa}, in the case of warped black holes.
For instance, we have done a similar work in Ref.\cite{Yekta:2015gja} for the minimal massive gravity theory\cite{Bergshoeff:2014pca} in the BTZ sector. 
\section*{Acknowledgment}
The author would like to thank A. Ghodsi for helpful comments.

\end{document}